# Towards Ultra-Large Bandwidth and a New Class of Specialty Optical Fibers

Sayan Bhattacherjee[1], AbhijitBiswas[1] and Somnath Ghosh[2*]

[1]Institute of Radio Physics and Electronics, University of Calcutta, Kolkata-700009, India
[2]Dept. of Physics, Indian Institute of Technology Jodhpur, Rajasthan-342011, India
*Corresponding author: somiit@rediffmail.com

We propose a scheme to enhance the effective photonic bandwidth exploiting bandgap overlapping of same order or different orders through judiciously chosen aperiodic geometries is spatial dimension. To implement the scheme, we design a specialty optical fiber with hybrid chirped-cladding. Our designed fiber provides an ultra-wide photonic bandwidth of ~ 3µm. The less dispersive behavior of the structure carries the signature of zero dispersion at ~2.7µm. The proposed two material all-solid fiber geometry is based on thermally compatible chalcogenide glasses, GeAsSe and AsSe as low and high index respectively. The efficient delivery of short pulses with less distortion, and enhanced frequency broadening of specific pulses at certain conditions are demonstrated through these specialty fibers.   © 2017 Chinese Laser Press

**OCIS codes:**(060.2310) Fiber optics;(060.5295) Photonic crystal fibers;(350.5500) Propagation.

## 1. INTRODUCTION

In late 1980 the communication through C-band flourished owing to discovery of Erbium Doped Fiber Amplifier (EDFA) **(1)**. The use of optical amplifiers though helped in the redundancy of electronic amplifiers as repeaters, the pulse dispersion effects would still be present as they accumulate with distance over multiple stages of amplifications. The signal pulses spread out in the time domain resulting in inter-symbol-interference and subsequently introduce bit errors in the data stream. To mitigate these problems various categories of these dispersion manipulating fibers such as dispersion shifted fibers (DSF), dispersion flattened fibers (DFF) **(2)** and dispersion compensating fibers (DCF) **(3)** were developed. Due to increasing demand of quality as well as bandwidth, there has been an urgent need for specialty fibers, in which transmission loss would not be limited by the material, together with the facility for tailoring of nonlinearity and dispersion properties to a greater degree of control. Here comes the band-gap guided fibers which offers itself as a potentially useful technology to mitigate many of such limitations.

The concept of photonic band-gap was introduced by simply suggesting a periodic stacking of regular array of channels of different dielectrics and drawing into waveguides/fiber form **(4)**. Later in 1987 Yablonovitch predicted the possibility of controlling light propagation in a crystal-like periodic medium by the photonic band-gap effect **(5)**, which had essentially led to a new paradigm in research on guided wave optics in the next generation of optical fibers. Thus in a band-gap engineered fiber, mechanism for guidance is totally different as compared to conventional TIR guided structures. It is because of the specialty in its structure where the refractive index of the central region is in general lower than the average index of the periodic cladding. Here Bragg scattered backward propagating light at each of the high and low index layer interface of the periodic cladding adds up in phase to build up light confinement in the low index central region, coined as core. The low index core region, which forms a localized defect due to breaking of the periodicity of the overall structure, can confine light of wavelengths within the photonic band-gap and allows propagation through it **(6)**. This phenomenon had opened up a new avenue with huge potential interest to meet many existing technological challenges. Thus various proposals were made in the field of defense, medical and advanced instruments. Lately, a novel design was proposed where a certain amount of aperiodicity was introduced across the cladding layers in their spatial and/or refractive index periodicity in the transverse plane which significantly helped modify fiber's characteristics as it provides more tuning parameters. **(7, 8, 9).**

In this paper, we propose a scheme to enhance the overall effective band-width for a chosen set of materials via overlapping the fundamental band-gap through the first higher order band-gap for the first time. Alternatively, different fundamental band-gaps of different periodic structures could be exploited to achieve ultra-wide PBG. We propose and report specially designed an all-solid Bragg fiber geometry consisted of suitable low-loss chalcogenide glasses with advanced cladding.This specialty fiber design can use the entire low-loss window of the material chosen and hence the overall loss is reduced over this bandwidth. The hybrid-chirped cladding photonic band-gap (PBG) fiber could be fabricated by well-known extrusion technique. The low-loss window of the designed fiber spans over 1.2µm to 4.2µm and the dispersion effects are also significantly reduced increasing the bandwidth upto 50% over conventional Bragg fibers. We have shown that our proposed fiber design could be used to transmit short optical pulses over a bandwidth of 3µm. This feature could be explored in bio-medical applications as well as in integrated optical devices. We have also studied the phenomenon of spectral broadening of high power optical pulses through such fibers. Due to high power self phase modulation (SPM) of the optical pulses takes place which results in generation of spectral broadening. We have exploited this phenomenon and tried to generate supercontinuum in mid-IR region. These new colors in mid-infrared have tremendous applications in gas sensing, environmental monitoring, industrial process control, and astronomy (10).

## 2. THEORY AND PROPOSED SCHEME

A Bragg fiber can be viewed as a solid cylindrical block covered by layers of various materials, one stacked over other. Fig 1(a) represents a schematic diagram for such geometry. The core region assumes the role of a defect in an otherwise periodic structure (11).

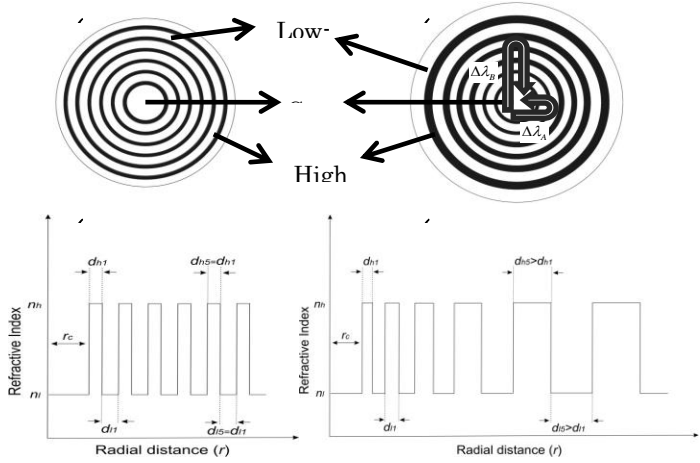

Fig. 1 (a) Schematic diagram of a Bragg fiber. (b) Schematic of refractive index profile of the Bragg fiber.(c)Schematic diagram of the proposed hybrid chirped-clad Bragg fiber supporting multiple band-gaps by cladding bi-layers of customized dimensions. (d) Schematic of refractive index profile of the hybrid chirped-clad Bragg fiber

The light of certain wavelengths are strongly confined through Bragg reflections by the surrounding specific multi-layers. These wavelengths are not discrete and random. The wavelengths which fall into the supported photonic band-gap of the structure are only strongly confined. At the band-edges there is a sharp increase of loss and dispersion (12). Lately, a special type of Bragg reflection waveguide (BRW) was proposed where the optical thickness of the cladding layers is such that round trip phase accumulated by light through one period (i.e., one pair of low and high refractive index cladding layers) equals $2\pi$, which is referred to as quarter wave stack condition for BRW (Qtw-BRW)(13, 11). Our proposed scheme evolves around the synthesis of two interesting concepts. The bi-layer thickness varies as we go away from core. This structure as shown in Fig 1(b) helps to enhance the band-gap because each bi-layer has certain band-gap (12). The bi-layers thicknesses are such chosen that there is continuous deviation from quarter wave stack condition. There is also a feature that each bi-layer supports discontinuous but multiple band-gaps. In our structure we have chosen the bi-layer thicknesses such that the discontinuity between fundamental and first higher order band-gap of last bi-layer vanishes. This is done by suitably chosen transverse chirp values such that there is continuous overlap of fundamental band-gaps. The schematic diagram of the refractive index profile of fibers are demonstrated by Fig. 1(c) and Fig. 1(d) During our analysis we have also found that the higher order band-gaps provide much tighter confinement but provide comparatively less bandwidth than fundamental band-gap (14). Therefore the same proposal of band-gap overlapping and tuning described above also applies for higher order band-gaps which help to enhance the bandwidth even further. Thus we get a much wider band-gap which will provide huge bandwidth in principle.

As a proof of concept of the proposed scheme, we design and optimize a specialty Bragg fiber with hybrid cladding. The contrastof the refractive indicesof the layers forming the advanced cladding chosen tobe such that $\Delta n \ll 1$. The low-index contrast of the cladding layers implies that the scalar linearly polarized (LP) approximation should be valid in these fibers. To analyze the propagation characteristics of these structures Bloch theorem is widely used, which is ideally valid for an infinitely extended periodic medium. However, in practice fibers have a finite cladding which makes the modes leaky in nature. Hence, the field in the last layer could be represented as a purely outgoing wave. This allowed us to analyze these Bragg fibers through a well known transverse matrix method (15) that was originally proposed for analyzing the leaky modes of a standard silica fiber. In this paper, we have adapted and extended this simple and efficient matrix method to study the modal characteristics of our fiber design.

According to this technique, the radial part of the modal field in various layers of the Bragg fiber under the *LP* approximation can be expressed in the following form:

$$R(r,z,\theta,t) = [A_j J_l(k_j r) + B_j Y_l(k_j r)] e^{i(\omega t - \beta z)} \begin{bmatrix} \cos(l'\theta) \\ \sin(l'\theta) \end{bmatrix}$$

j = 1,2,3…N ;   $l, l' = 0,1,2…$

where $k_j^2 = (n_j k_0)^2 - \beta^2$, $\beta$ is the propagation constant, $k_0$ is the free-space wave-vector, $n_j$ is the refractive index of the jth region and N is the total number of cladding layers.

Study of evolution of optical pulses through the lowest order modes in the designed fiber is modelled by solving the following nonlinear Schrödinger equation (NLSE) through the symmetrised split-step Fourier method:

$$\frac{\partial A}{\partial z} + \beta_1 \frac{\partial A}{\partial T} + i\frac{\beta_2}{2}\frac{\partial^2 A}{\partial T^2} - \frac{\beta_3}{6}\frac{\partial^3 A}{\partial T^3} + \frac{\alpha}{2}A = i\gamma(\omega_0)|A|^2 A$$

where, nonlinear parameter $\gamma$ is defined as

$$\gamma(\omega_0) = \frac{n_2(\omega_0)\omega_0}{cA_{eff}}$$

α is the loss parameter, $\beta_1$ is the first dispersion, $\beta_2$ second order dispersions (GVD) and$\beta_3$ is the third order dispersion (TOD).$A_{eff}$ is the effective mode area of the mode and $n_2$ is the nonlinear coefficient of the material. The pulse amplitude is assumed to be normalized such that $|A|^2$ represents the optical power. We have used pico-second pulses for our analysis. For this reason we have not taken into account the terms which include cross phase modulation, Raman and Brillion gain effects. These effects can be neglected for picoseconds pulses but are unavoidable for sub pico-second pulses.

## 3. FIBER DESIGN AND PERFORMANCES

To justify the efficiency of our proposed scheme, we report a quantitative analysis of the performance of a conventional Bragg fiber against our proposed hybrid-chirped cladding fiber. Both the fibers for the analysis, have been designed to have 6 bi-layers. To assure single mode guidance, the core of the fiber is assumed to have diameter of 7 µm. In case of Bragg fiber, all the bi-layer dimensions are fixed such that the thicknesses of high-index layers ($d_h$= 0.6µm) and thicknesses of low-index layers ($d_l$ = 1.5µm) respectively. However, for our designed hybrid-chirped cladding fiber we have aperiodically varied thicknesses of high-index layers from 0.6 µm to 0.9 µm and thickness of low-index layers from 1.5 µm to 2.7 µm over 6-bilayers. Details of the bi-layer dimensions of the hybrid-chirped cladding are given in Table 1 given below.

Table 1. Details of bi-layer thickness

| Bi-layer Number | Thickness of high-index bi-layer ($d_h$) in µm | Thickness of low-index bi-layer ($d_l$) in µm |
|---|---|---|
| 1 | 0.6 | 1.5 |
| 2 | 0.61 | 1.53 |
| 3 | 0.65 | 1.62 |
| 4 | 0.66 | 1.65 |
| 5 | 0.82 | 2.45 |
| 6 | 0.9 | 2.7 |

We have proposed the fiber design consisting of two types of low-loss chalcogenide glasses, each doped with such a way that their refractive indices vary by a small value $\Delta n \ll 1$. To ensure the feasibility of fabrication of the fiber, we have used a pair of thermally compatible chalcogenide glasses. To calculate the refractive indices of chalcogenide glasses at different wavelengths we have used the following formula:

$$n = \sqrt{(1+((B_1*\lambda^2)/(\lambda^2-C_1^2))+((B_2*\lambda^2)/(\lambda^2-C_2^2))+((B_3*\lambda^2)/(\lambda^2-C_3^2)))}$$

η is the refractive index of the material at different wavelength λ. $B_1$, $C_1$, $B_2$, $C_2$, $B_3$ and $C_3$ are the specific coefficient for each glass. Theses coefficients are tabulated in Table 2.

Table 2. Details of coefficients of chalcogenide glasses

| Material Coefficient | For high-index (AsSe) chalcogenide glass | For low-index (GeAsSe) chalcogenide glass |
|---|---|---|
| $B_1$ | 3.1918 | 2.8498 |
| $C_1$ | 0 | 0 |
| $B_2$ | 3.5884 | 3.0044 |
| $C_2$ | 0.4306 | 0.4000 |
| $B_3$ | 2.4035 | 1.1265 |
| $C_3$ | 60.3285 | 47.4294 |

To show that our fiber design scheme is indeed extremely versatile and useful in enhancement of bandwidth, we have done performance analysis for our designed fiber against a conventional Bragg fiber counterpart. This is achieved by generating band-gap plots as well as by calculating of low-loss window offered by the structures. The band-gap plot shown in Fig. 2 justifies that due to the customized advanced cladding, effective band-gap is significantly enhanced. In this figure the band diagram due to first bi-layer is shown in blue color. Now if the bi-layer dimension across the finite cladding structure remains the same, there is no enhancement of band-gap as evident as proposed. Hence we have designed and exploited an advanced cladding structure. Band structure of the last bi-layer of this advanced cladding is shown in the figure by red color.

Therefore it can be visualized that if we could merge the fundamental band-gaps of innermost and outermost bi-layers, then one would certainly open-up a much wider band-gap. To appreciate this, the bi-layer dimensions in-between first and last bi-layer are judiciously chosen and optimized. An enlarged view of a selected portion of the band diagram of Fig. 2a is shown in Fig. 2b.

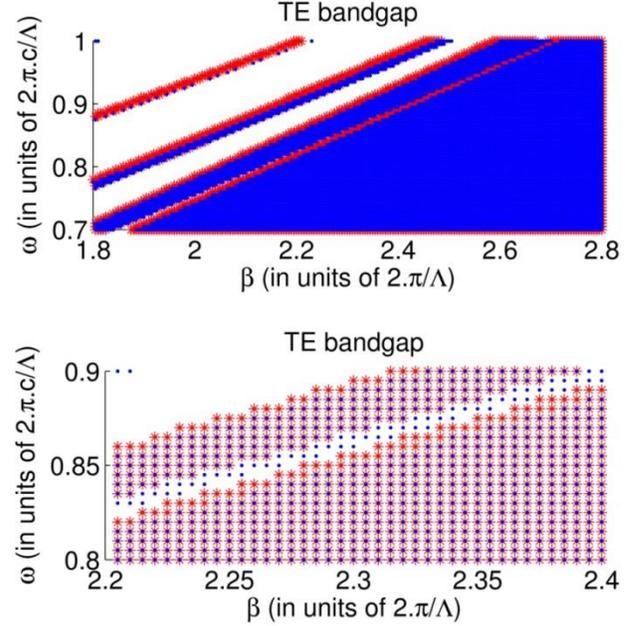

Fig. 2 (a) Band diagrams (TE) of two ideal Bragg fibers having periodic claddings; first (shown in blue dots), for which $d_h$ = 0.6 µm, $d_l$ = 1.5 µm, and the second (shown in red stars), for which $d_h$= 0.9µm, $d_l$ = 2.7µm. In both the cases, we have assumed the periodicity is invariant over infinite cladding. (b) Zoomed view over small portion of Fig. 2a.

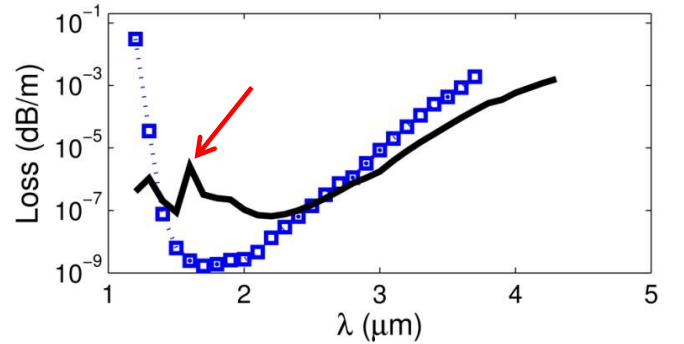

Fig. 3 Numerically simulated low-loss window of Bragg fiber (shown by blue dotted line) and its counterpart hybrid-chirped clad fiber (shown by black continuous line).

The low-loss windows of the two different fiber geometries are shown in Fig. 3. From a direct comparison, it can be observed that there is an enhancement of low-loss window in case of hybrid-chirped cladding fiber. For Bragg fiber the effective window is from ~1.2 µm to ~3.2 µm; but for hybrid-chirped cladding fiber the window is from ~1.2 µm to ~4.2 µm. This claim is well established when the dispersion curve shown later is analyzed. Thus the estimated enhancement factor in terms of low-loss window came out to be more than 50% for our fiber design. Further it can be observed from the figure that though in the case of Bragg fiber minimum of the low-loss window, achieved in comparison to our fiber design is less; however for practical purposes we require an almost flat transmission window so that amplifiers can be explored over a wider window. The small peak around 1.6µm shown by red arrow confirms the overlap of two well separated band-gaps. This can be understood by synthesis of two separate phenomena, first is much tighter confinement is achieved by higher order band-gaps and second is the fact that the central wavelength of fundamental band-gap is proportional to the dimension of the bi-layers. Also it should be kept in mind that higher order band-gaps occur at lower wavelength than fundamental band-gap. During designing the fiber we have taken into account these features. Thus each wavelength is not confined by same set of bi-layers. The peak shown by red arrow is a region where the wavelengths are confined by less than 6 bi-layers, because all the 6 bi-layer do not have band-gap in this region as bi-layer dimensions are constantly and discretely varying. The peak also carries the signature of merging fundamental as well as higher order band-gap. Moreover, choice of more than 6 bi-layers will wash out presence of any such unwanted kinks, however the total operational bandwidth would reduce as the bi-layers far away from the core would not have any effect on the over-all guiding mechanism through the core. Additionally, by customizing the cladding dimensions within only 6 bi-layers reduces the cost of fabrication. We have also calculated and plotted the mode line of our fiber design (shown by red line) around 1.6 µm on the band diagram as shown in Fig. 4. This justifies that we are able to merge the fundamental as well as higher order band-gap by designing an advanced hybrid-chirped cladding.

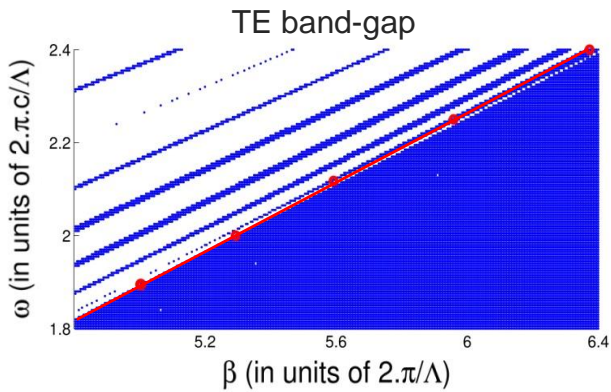

Fig. 4 Mode line (shown in red color) plotted over band diagrams of ideal Bragg fibers having periodic claddings for which $d_h$ = 0.9µm, $d_l$ = 2.7µm.

One can appreciate the enhancement of band-gap and hence the low-loss window from the aforementioned results. But mere enhancement of low-loss window does not confirm the enhancement of transmission bandwidth. To establish a wideband transmission window the dispersion characteristic of the fiber should also be studied. Highly dispersive PBG structures lead to pulse spreading and essentially deleterious signal distortion/information loss. However, if the dispersion values are almost negligible, then it leads to four wave mixing (FWM) causing unnecessary optical kinks over the desired spectra of operation. Thus to design an efficient transmission window we should design the fiber such that the dispersion behavior is flattened enough over the desired window of operation. To ensure this type of dispersion characteristics, the effect on dispersion values due to overlapping widths of band-gaps is investigated. As mentioned in the previous section the dispersion values increases as we move towards the edges of the band-gap window. Straight forwardly, to obtain a flat dispersion curve we should have a wider band-gap. However, for wider band-gaps we need thicker bi-layers and selection of wavelength of operation is dictated by this bi-layer thickness. Thus a flat dispersion curve at smaller wavelength with enhanced over-all bandwidth cannot be achieved using any ideal Bragg fiber. The advantage of customized hybrid-chirped cladding fiber over the ideal Bragg fiber is that it provides a continuous wide band-gap. Thus it can provide a flat dispersion curve over much wider window. As the dispersion curve is flat for hybrid-chirped cladding fiber the group velocity dispersion (GVD) effects of our fiber design become tremendously improved in comparison to an ideal Bragg fiber. Findings of these characteristics is displayed in Fig. 5(a) and Fig. 5(b) respectively.

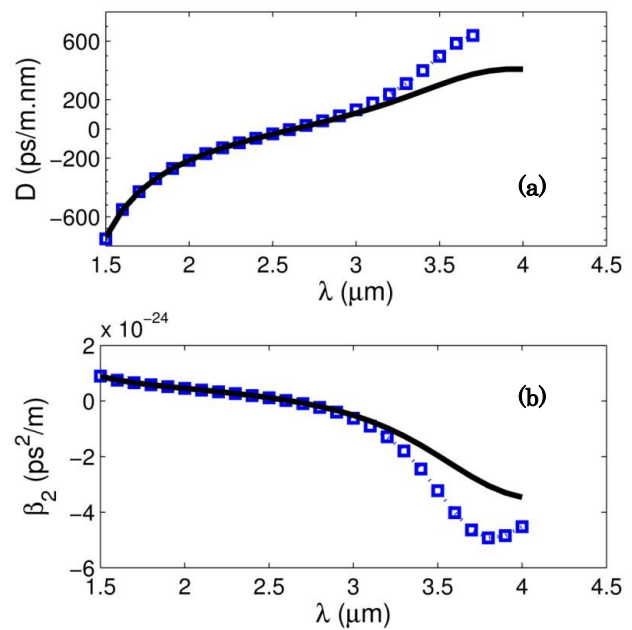

Fig. 5(a) Numerically simulated dispersion curve of Bragg fiber (shown by blue dotted line) and designed hybrid-chirped cladding fiber (shown by Black continuous line). (b) Numerically simulated GVD parameter of Bragg fiber (shown by blue dotted line) and designed hybrid-chirped cladding fiber (shown by black continuous line).

As the dispersion behavior over wider wavelength range is the key to control pulse propagation, in the next section we have exploited the evolution of pulse propagating through our fiber design. The promising features of our fiber design are ultra-flat dispersion and GVD characteristics which can be exploited for optical communication applications and bio-medical pulse delivery for ablation and other therapeutic applications. On the other hand, the characteristics obtained by our fiber design are very much suitable for generating efficient spectral broadening for specific cases over a short length of propagation using high power optical pulse as input. The characteristics of the generated continuum spectra can be improved by parametric control.

## 4. EVOLUTION OF OPTICAL PULSE

Split step Fourier method is a very efficient numerical approach for analysing the nonlinear behavior of propagating pulses in optical fibers (16). The calculation is done in two steps. The ansatz follows as, during propagation of optical field over a small distance "*h*", the dispersive and non-linear effects act independently. Over the first step the non-linearity acts alone and dispersive effect is ignored; whereas over the second step its vice-versa. The appropriately chosen small distance over which this calculation is done is a fraction of the non-linear length or dispersive length of the fiber (whichever is less). Thus the prerequisite of solving split step Fourier method is calculating the dispersive and nonlinear linear length:

$$L_{NL} = \frac{1}{\gamma P_0} \quad ; \quad L_D = \frac{T_0^2}{|\beta_2|}$$

Where, $L_{NL}$ = Non-linear length; $L_D$ = Dispersive length;
$P_0$ = Peak power; $T_0$ = Time period of pulse;
$\gamma$ = Non-linearity parameter; $\beta_2$ = GVD of fiber

### Spectral Broadening

To demonstrate the phenomenon of spectral broadening, we have to study the evolution of optical pulse through the fibers such that SPM/dispersive broadening take place. To ensure the occurrence of SPM we have assumed the input pulse to be Gaussian in nature with a peak power of 1KW and pulse width of 1ps. We have simulated for optical pulse centered at 3.1 µm as dispersive nature of our fiber design is improved over the conventional Bragg fiber at this wavelength. It can be clearly observed from Fig. 6 that the spectral broadening through our designed hybrid chirped cladding fiber is wider compared to the conventional Bragg fiber after propagating through a certain distance.

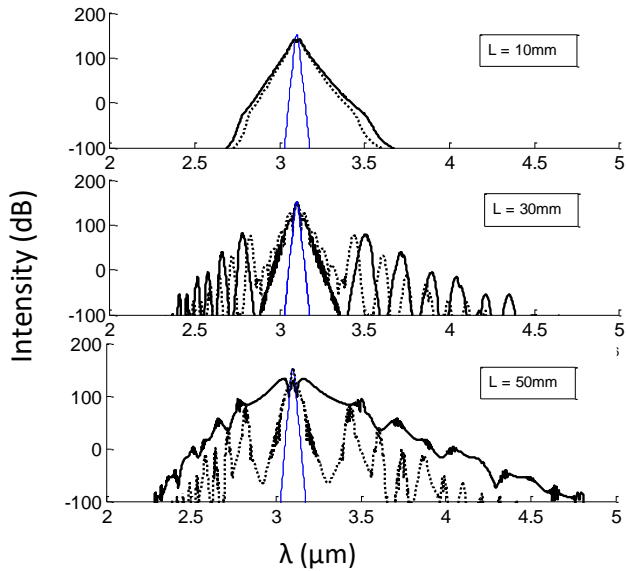

Fig. 6 Evolution of a Gaussian input pulse is of peak power 1KW and pulse width 1ps, on travelling through the fiber of various lengths (shown in the boxes). The solid line is for hybrid chirped cladding fiber and dotted line is for conventional Bragg fiber.

To understand the characteristic feature of much broader spectra for hybrid chirped fiber in comparison to Bragg fiber, we looked into several aspects. The non-linear length of the hybrid chirped fiber is less compared to the Bragg fiber. As a result new frequencies would be generated which would contribute to broaden the spectra. We have also simulated for input Gaussian pulse of peak power of 10KW and duration of 1ps. The comparison of output pulse is shown in Fig. 7. As expected the spectrum is much wider as the phenomenon of SPM occurs favourably at much smaller length due to high power and new frequencies are generated which again lead to successive generation of new frequencies and hence enhancing the phenomenon of spectral broadening. Effect of third order dispersion has been deliberately ignored here.

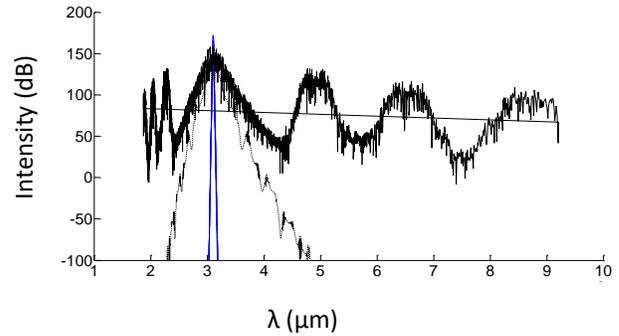

Fig. 7 Evolution of a Gaussian input pulse of 1ps duration through hybrid chirped cladding fiber of 50 mm length. The solid line is for peak power of 10KW and dotted line is for peak power of 1KW

We study propagation of short optical pulses through the designed hybrid chirped clad fiber and discuss the advantages of in context of specific applications of short undistorted pulse delivery over the counterpart conventional Bragg fiber.

A direct comparative analysis of the characteristic length scales of dispersion and nonlinear behavior of the fibers are taken into account at 3.1 µm and 4.1 µm respectively. This specific choice of wavelengths which are 1 µm apart in the loss window are to appreciate the enhancement of bandwidth due to hybrid chirped cladding fiber over a conventional Bragg fiber. The peak power of input optical pulse used for simulation is kept low such that there is no SPM of the optical pulse in this context. During our analysis at 3.1 µm, we have assumed the peak power and width of input Gaussian pulse to be 20W and 2ps respectively. Similarly for analysis at 4.1 µm we have assumed the peak power and width of input Gaussian pulse to be 20W and 10ps respectively.

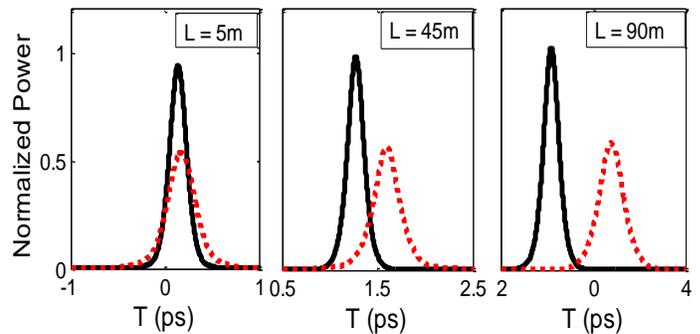

Fig. 8 Evolution of a Gaussian input pulse of 20W peak power and 2ps duration through hybrid chirped cladding fiber (shown by black solid line) and that of a conventional Bragg fiber of various lengths (shown by red dotted line) at 3.1 µm.

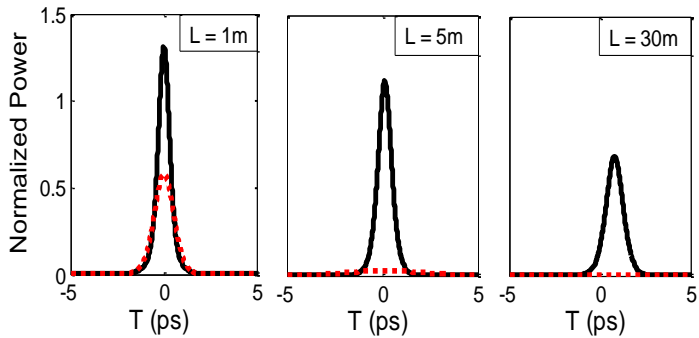

Fig. 9 Evolution of a Gaussian input pulse of 20W peak power and 10ps duration through hybrid chirped cladding fiber (shown by black solid line) and that of a conventional Bragg fiber of various lengths (shown by red dotted line) at 4.1 µm

From the Fig. 8 and Fig. 9 it is evident that the hybrid chirped cladding fiber design is much more efficient than the conventional Bragg fiber to deliver stable undistorted pulses over longer distances. At 3.1 µm pulse can be transmitted through both the fiber but the accumulated phase shift encountered by pulse through the Bragg fiber is much higher. Whereas in the case of transmission at 4.1 µm, the optical pulse through the Bragg fiber hardly reaches upto 5m in comparison to hybrid chirped cladding fiber where stable pulses are reported upto 30m distance. Thus hybrid chirped cladding fiber has a new degree of freedom in enhancing the bandwidth of operation as well as propagation distance.

## 5. SUMMARY

In summary, our proposed scheme of bandgap overlapping and design of hybrid chirped cladding fiber by tailoring the quasi-periodicity in an otherwise periodic PBG structure has been proven to be an efficient tool for customizing the characteristics of specialty optical fiber. We have established the advantages of our design for ultra wide-band transmission exploiting its flat low-loss and less-dispersion characteristics. This feature can also be exploited in bio-medical application for short pulse delivery and real time control. Moreover, we have also shown that our designed fiber can be potentially used to generate new wavelengths at mid-IR range to open up potential applications in the field of sensing, imaging as well as astronomy.

**Funding.** SG acknowledges financial support from Department of Science and Technology, India [IFA-12; PH-13].